\documentclass[aps,prb,twocolumn,showpacs,twoside,amsmath,amsfonts]{revtex4}
\usepackage{graphicx}
\usepackage{epsfig}
\usepackage{amssymb}
\begin{document}
\title{Multiphoton Processes in Driven Mesoscopic Systems. }
\author{Alessandro Silva}
\affiliation{The Abdus Salam International
Centre for Theoretical Physics, Strada Costiera 11, 34100 Trieste,
Italy}
\author{Vladimir E. Kravtsov}
\affiliation{The Abdus Salam International Centre for Theoretical
Physics, Strada Costiera 11, 34100 Trieste, Italy}
\affiliation{Landau Institute for Theoretical Physics, 2 Kosygina
st., 117940 Moscow, Russia}
\begin{abstract}
We study the statistics of multi-photon absorption/emission
processes in a mesoscopic ring threaded by an harmonic
time-dependent flux $\Phi(t)$. For this sake, we demonstrate a
useful analogy between the Keldysh quantum kinetic equation for the
electrons distribution function and a Continuous Time Random Walk in
energy space with corrections due to interference effects. Studying
the probability to absorb/emit $n$ quanta $\hbar\omega$ per
scattering event, we explore the crossover between
ultra-quantum/low-intensity limit and quasi-classical/high-intensity
regime, and the role of multiphoton processes in driving it.
\end{abstract}
 \maketitle


\section{Introduction}

Recently a surge of interest in the dynamical properties of
mesoscopic/nanoscale electronic systems has motivated a number of
theoretical and experimental studies on the physics of electronic
devices subject to the driving of external fields. This main theme,
pioneered in Ref.~\onlinecite{Falko}, embraces a number of
interesting issues such as the study of the influence of microwave
driving on transport through chaotic scatterers~\cite{AlVav}, the
phenomenon of adiabatic quantum pumping~\cite{pump}, as well as
diffusion and localization~\cite{diffuzion1,diffusion2,BSK,Houches}
in energy space in quantum chaotic systems/disordered quantum dots.
In a broader context, the effect of the driving of external
microwave fields has been shown to lead to an intriguing zero
resistance state in quantum Hall systems~\cite{QH}, and is currently
studied as a tool to control the coherent dynamics of
superconducting Josephson qubits~\cite{Qubit}.

Investigations of periodically driven mesoscopic systems/quantum
dots addressed mostly the limit of low intensity
driving~\cite{AlVav,BSK,Houches}. In this case, electrons have
enough time to explore ergodically all available phase space before
performing a single photon assisted transition in energy space. This
makes it possible to use an effective time dependent Random Matrix
Theory to describe the dynamics of the system. On the other hand, as
beautifully shown by recent experiments in superconducting
qubits~\cite{Qubit} and in the quantum Hall regime~\cite{QH}, as the
intensity of driving increases one should expect both an enhancement
of the probability of single-photon processes, and the emergence of
multi-photon processes/resonances in the dynamical properties of the
system under study.

{\setlength{\unitlength}{1cm}
\begin{figure}
\vglue 2cm \epsfxsize=0.7\hsize
\hspace{-3cm}
\begin{picture}(4,4)
   \epsffile{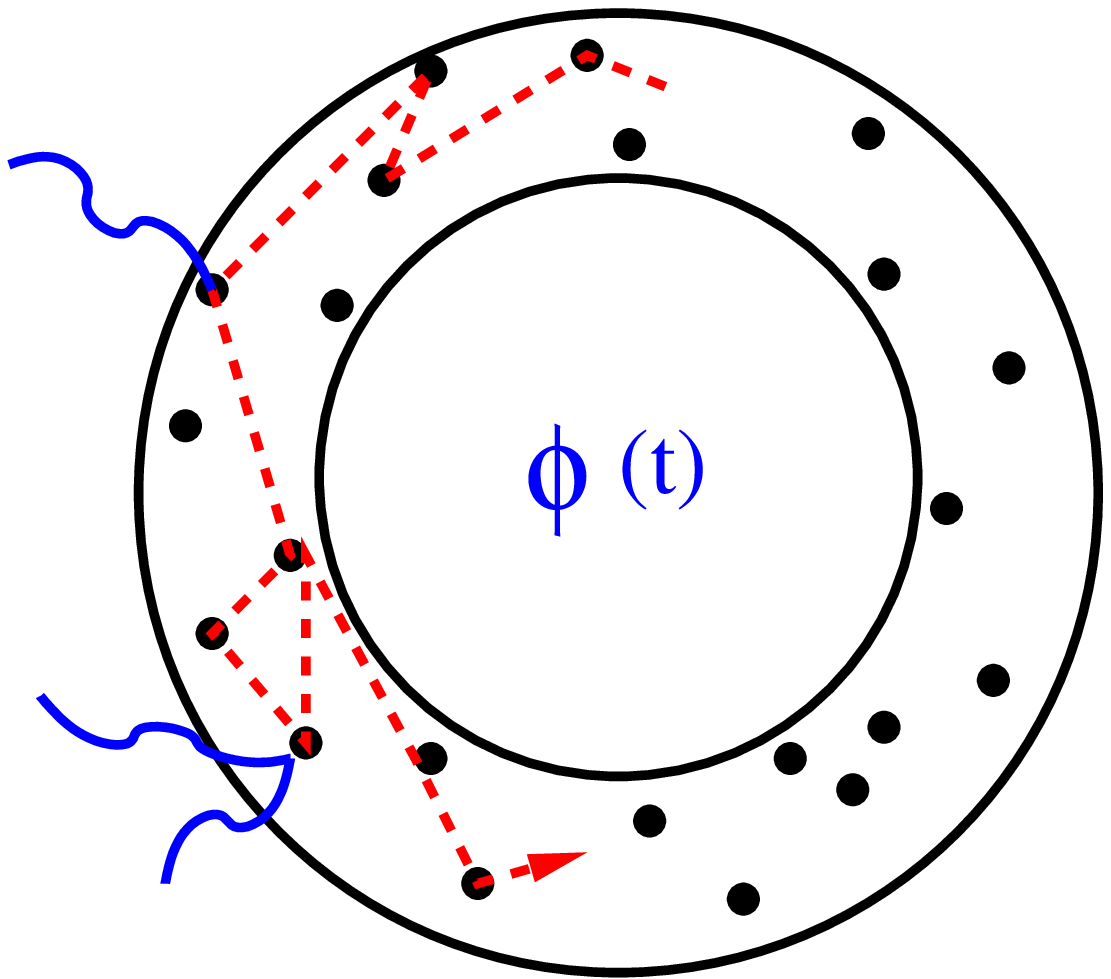}
\end{picture}\vspace{0cm}
\vspace{1cm} \caption{The physical system under study, a diffusive
quasi one dimensional ring thread by a time dependent flux
$\Phi(t)$. For an harmonic time dependence
$\Phi(t)=\bar{\Phi}\cos(\omega t)$, the scattering of electrons off
impurities induces transitions in energy space quantized in units
$\hbar \omega$. The statistics of such transitions and its physical
consequences as the intensity of driving grows  are described by
Eq.(\ref{KE}) and Eq.(\ref{HG}).} \label{Fig0}
\end{figure}}

The goal of this paper is to characterize the influence of
multiphoton processes on the dynamical properties of mesoscopic
electronic systems, concentrating on their effect on the electron
dynamics in energy space (diffusion/localization). Diffusion and
localization in energy space, as well as of multi-photon processes,
have been the subject of a number of studies in the context of the
optics of complex atoms/molecules~\cite{Akulin}. In these systems
the underlying electron dynamics is typically very complex and a
statistical description, either equivalent to random matrix theory
or explicitly using it, is compulsory. In contrast, in the present
study we go beyond random matrix theory focusing on a model
mesoscopic system, a diffusive quasi-one dimensional ring threaded
by an oscillating flux (see Fig.\ref{Fig0} and Eq.(\ref{Ham})), were
the underlying microscopic dynamics can be studied in detail. We
explore how multi-photon processes proliferate as the driving
amplitude is increased or the driving frequency is decreased, study
the resulting crossover from ultra-quantum/low-intensity limit to
quasi-classical/high-intensity limit, and extract its physical
consequences. On the theoretical side, we demonstrate and use
extensively an interesting analogy between the quantum kinetic
equation for the electron distribution function and the recursion
relation defining a Continuous Time Random Walk~\cite{CTRW} in
energy space.

The rest of the paper is organized as follows. In
Sec.~\ref{qualitative} we present qualitatively the results of our
analysis of diffusion in energy space and of multiphoton processes
based on the mapping of the problem onto a continuous time random
walk in energy space. This mapping is derived in full detail in
Sec.~\ref{quantitative} using the Keldysh technique. Finally, in
Sec.~\ref{conclusions} we present our conclusions.


\section{Qualitative Analysis}~\label{qualitative}

In this section we start by summarizing the qualitative picture
emerging from our analysis. The elementary time scale controlling
the dynamics of energy absorbtion/emission is the mean free time
$\tau$. Indeed, in a diffusive quasi-1d ring threaded by a flux
$\Phi(t)=\bar{\Phi} \cos(\omega t)$, energy changes quantized in
units $\hbar \omega$ occur provided an electron scatters off an
impurity. This is due to the fact that during the ballistic
trajectory in between scattering events the flux perturbation
$V(t)=-A(t)\,\hat{v}$, $\hat{v}$ being the velocity operator,
commutes with the unperturbed Hamiltonian $H_{0}=m \hat{v}^{2}/2$,
therefore causing no transitions whatsoever.

In the \it ultra-quantum \rm limit of weak perturbations
single-photon processes dominate. In other words, in one scattering
event an electron may either absorb/emit one quantum, or scatter
elastically. In particular, the probability $P_{\Omega}$ to make a
transition of energy $\Omega$ in energy space in one scattering
event is given by
\begin{equation}
\label{RMT} P_{\Omega}=(1-p)\delta(\Omega)+\frac{p}{2}\,
\left[\delta(\Omega-\hbar\omega)+\delta(\Omega+\hbar\omega) \right],
\end{equation}
where $p \propto (\bar{\Phi}/\Phi_0)^2 \ll 1$, $\Phi_0$ being the
flux quantum.

On the other hand, in the opposite limit of high intensities of the
perturbation it is natural to expect {\it quasi-classical}
continuous energy absorption described by a Drude-like picture.
According to this picture an electron moving ballistically between
two scattering events (at times $t'$ and $t$, respectively) acquires
an energy $\int_{t'}^{t}dt''\, e v_F \vec{E}(t'')\cdot \hat{n}$,
where $\vec{E}(t)=-\partial_t \vec{A}(t)$ is the electric field
generated by the time dependent flux, and $\hat{n}$ is the momentum
direction in the $d$-dimensional space. Let us introduce the
probability density ${\cal P}_{\Omega}(t,t')$ of changing the energy
by $\Omega$ between two successive scattering events at $t'$ and
$t$. Given the Poisson distribution
\begin{eqnarray}\label{Poisson}
\psi(t-t')=\frac{1}{\tau}\;e^{-|t-t'|/\tau},
\end{eqnarray}
of time intervals $|t-t'|$, neglecting acceleration by an electric
field $\vec{E}(t)$, and assuming isotropic scattering, one may
immediately write
\begin{eqnarray}\label{clabs} {\cal P}_{\Omega}(t,t')=\psi(t-t')
\left\langle \delta\left(\Omega -\int_{t'}^{t}dt'' e v_F
\vec{E}(t'')\cdot \hat{n}\right) \right\rangle_{\hat{n}}
\end{eqnarray}
where $\langle \ast \rangle \equiv \int d\hat{n}\;(\ast)$ denotes
the averaging over momentum directions.
For an harmonic time dependence $E(t)=E_{0}\,\cos(\omega t)$, and at
low frequencies $\omega\tau\ll 1$, one obtains
\begin{equation}
\label{P-w} P_{\Omega}=\int_{-\infty}^{t}\langle{\cal
P}_{\Omega}(t,t')\rangle_{T}\,dt'=\biggl\{\begin{matrix}
\langle\frac{E_{1}(\frac{|\Omega|}{\Omega_{0}\sin(\omega t
)})}{2\Omega_{0}\sin(\omega t)}\rangle_{T},& \!3d\! \cr
\langle\frac{K_{0}(\frac{|\Omega|}{\Omega_{0}\sin(\omega t
)})}{\pi\Omega_{0}\sin(\omega t)}\rangle_{T}, & \!2d\! \cr
\end{matrix}
\end{equation}
where $\langle * \rangle_{T}=
\int_{-\frac{\pi}{\omega}}^{\frac{\pi}{\omega}}\frac{dt\omega}{2\pi}(*)$
denotes averaging over the period,
$E_{1}(z)=\int_{1}^{\infty}e^{-zt}\,\frac{dt}{t}$ is the exponential
integral function, $K_{0}(z)$ is the Bessel function, and
$\Omega_{0}=eE_{0}v_{F}\tau$.

Multiphoton processes drive the crossover between discrete energy
absorbtion in the ultra-quantum limit [Eq.(\ref{RMT})], and
continuous energy absorbtion in the quasi-classical limit
[Eq.(\ref{P-w})]. As shown below, the crossover probability function
$P_{\Omega}$ may be written as $P_{\Omega}=\sum_n \:P_n
\delta(\Omega-\hbar n\omega)$. In the low frequency limit $\omega
\tau \ll 1$, and for isotropic 3d scattering, we obtain
\begin{eqnarray}
\label{HG} P_n = {\cal E}^{2n} A_n\; _{3}{\cal F}_2\left[
{a}_n,{b}_n,-16{\cal E}^2\right]
\end{eqnarray}
where ${\cal E}=eE_{0}v_{F}\tau/(\hbar\omega)$, $_{3}{\cal F}_2$ is
a generalized hypergeometric function, $A_n=
(2^{2n}\Gamma[n+1/2])/(\sqrt{\pi}(1+2n)\Gamma[n+1])$,
${a}_n=\{n+1/2,n+1/2,n+1/2\}$, and ${b}_n=\{n+3/2,1+2n\}$. These
functions are plotted for selected values of $n$ in Fig.1.

At low intensities (${\cal E} \ll 1$) the probability to absorb/emit
$n$ photons is
\begin{equation}
\label{smallE} P_n =A_{n}\, {\cal E}^{2n} \;\;\;\; {\cal E }\ll 1.
\end{equation}
Therefore, multi-photon processes are exponentially suppressed.
Neglecting them we obtain Eq.(\ref{RMT}) with $p={\cal E}^{2}/6$. As
the intensity grows, higher order processes become increasingly
probable at the expense of single (or in general low) order ones, as
indicated by the fact that for ${\cal E} > 2$, $P_1({\cal E})$
starts decreasing. At ${\cal E}\gg 1$ Eq.(\ref{HG}) can be
approximated as follows:
\begin{equation}
\label{Elar} P_{n}\propto\frac{1}{{\cal E
}}\left\{\begin{matrix}\ln^{2}({\cal E}/n), & {\cal E}\gg n \cr
\exp[-n/{\cal E}], & {\cal E}\ll n\cr
\end{matrix} \right.
\end{equation}
Note that  in the interval $1\ll n\ll {\cal E}$ the probability of
absorbing/emitting $n$ photons decrease very slowly with increasing
$n$ which leads to a proliferation of multi-photon processes at
large ${\cal E}$.

The two distinct regimes of rare (${\cal E} \ll 1$) and
proliferating (${\cal E} \gg 1$) multiple photon processes have been
discussed by Keldysh~\cite{MultiKeldish} in his seminal work on atom
ionization. In this case the two regimes are classified by the ratio
$\gamma^{-1}=\omega_t/\omega$ of the inverse time
$\omega_t=eE_0/\sqrt{mI_0}$ of tunneling through the potential
barrier titled by the electric field $E_0$, $I_0 \gg \hbar\omega$
being the ionization threshold, and of the frequency of the applied
electromagnetic field $\omega$. The qualitative connection between
Ref.[\onlinecite{MultiKeldish}] and our problem is obtained by
identifying $\gamma^{-1}$ with ${\cal E}$, and $I_0$ with $1/\tau$
in the present analysis.

\begin{figure}
   \epsffile{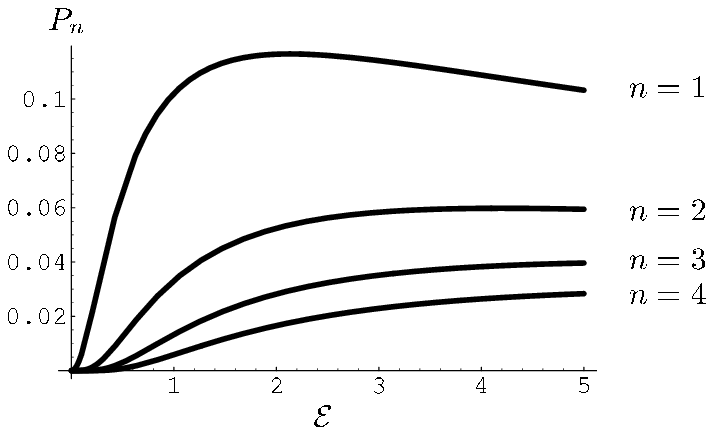}
   \vspace{0.5cm}
\caption{The probability to absorb/emit $n$ photons in a scattering
event $P_n$ plotted as a function of the intensity parameter ${\cal
E}$ for selected values of $n$. At low intensities, single photon
processes dominate. At higher intensities, higher order multiphoton
processes become increasingly important, driving a
quantum-to-classical crossover [see text].} \label{Fig1}
\end{figure}

Let us now make the qualitative considerations above more precise.
As shown in Sec.\ref{quantitative} in terms of a Keldysh
diagrammatic analysis~\cite{Houches,Keldysh, YKK}, the dynamics of
energy absorption/emission in the system at hand may be conveniently
described as {\it random walk} in the energy space described by the
recursion relation for the electron energy distribution function
$f_{t}(E)$
\begin{eqnarray}\label{KE}
f_t(E)=  \int_0^{t} dt' \int_{-\infty}^{+\infty} d\Omega\; {\cal
P}_{\Omega}(t,t')\; f_{t'}(E-\Omega).
\end{eqnarray}
Neglecting weak localization effects controlled by the parameter
$\lambda_f /v_F \tau\ll 1$ and effects of dynamic localization
controlled by the parameter $\delta/(v_F^{2}\tau A(t)^{2})\ll 1$
(where $\delta$ is the mean separation of electron levels in a
finite system)~\cite{BSK}, the kernel ${\cal P}_{\Omega}$, is given
by the product of two functions ${\cal P}_{\Omega}(t,t')=\psi(t-t')
p_{\Omega}(t,t')$. The function $\psi(t-t')$, given by
Eq.(\ref{Poisson}), is the distribution of the ballistic time of
flight $|t-t'|$, which may be interpreted as the continuous waiting
time in between steps of a random walk in the energy space. The
other function, $p_{\Omega}(t,t')$, is the conditional probability
to absorb/emit an energy $\Omega$ during the ballistic flight and is
given by $p_{\Omega}(t,t')=\int
d\eta\,e^{-i\Omega\eta}\,\tilde{p}_{\eta}(t,t')$, where
\begin{eqnarray}
\tilde{p}_{\eta}(t,t')&=& \left\langle e^{i v_F \hat{n}\cdot
\left[(\int_{t'+\eta/2}^{t+\eta/2} -\int_{t'-\eta/2}^{t-\eta/2})
dt'' e \vec{A}(t'')\right] }\right\rangle_{\hat{n}}\label{quantum}.
\end{eqnarray}
This result holds for a generic time dependence of
$\vec{A}(t)=\hat{n}_{x}\Phi(t)/L$ as long as $e|\vec{A}(t)|\lambda_F
\ll 1$. A random walk  of the type defined by the recursion relation
Eq.(\ref{KE}) is known in literature as a {\it Continuous Time
Random Walk} (CTRW) \cite{CTRW}.

One may now easily derive the crossover probability function
Eq.(\ref{HG}). Indeed, in the case of harmonic flux
$A(t)=\frac{E_{0}}{\omega}\,\cos(\omega t)$, one obtains
\begin{eqnarray}\label{discrete} p_{\Omega}(t,t')&=
&\sum_{n=-\infty}^{+\infty} \delta(\Omega -n\hbar\omega)
\;p_n(t,t')
\\ p_n(t,t')&=&\left\langle J_{2n}\left[ 2\frac{{\cal E}}{\omega
\tau} \hat{n}\cdot \hat{n}_x \left[ \cos(\omega t)-\cos(\omega t')
\right]\nonumber \right]\right\rangle_{\hat{n}},
\end{eqnarray}
where $J_{2n}(z)$ is a Bessel function and ${\cal E}=eE_{0}v_F
\tau/\hbar\omega$.
Though $p_{n}(t,t')$ does not explicitly depend on $\tau$, the
probability (averaged over the period $T$ of flux oscillations)
\begin{equation}
\label{P-n} P_{n}=\int_{-\infty}^{0}dt'\,\psi(t-t')\,\langle
p_{n}(t,t') \rangle_{T},
\end{equation}
of a multi-photon process between two successive scattering events
does depend on the mean free time $\tau$ through the function
$\psi(t)$. For $\omega\tau\ll 1$ one expands the difference of
cosines in Eq.(\ref{discrete}), introduce a new variable
$(t-t')/\tau$ and immediately concludes that $P_{n}\equiv
P_{n}({\cal E})$ is a function of ${\cal
E}=eE_{0}v_{F}\tau/(\hbar\omega)$. Performing explicit integrations
in $3d$ leads finally to Eq.(\ref{HG}). Similar results may be
derived in the high-frequency limit $\omega\tau\gg 1$. In this case
where one may set $\psi(t-t')\approx 1$ and observe that
$P_{n}\equiv P_{n}(\tilde{{\cal E}})$ is a function of the
$\tau$-independent parameter $\tilde{\cal
E}=eE_{0}v_{F}/\hbar\omega^{2}$.


It is now possible to show directly that the discrete probability
distribution $P_n$ interpolates between \it ultra-quantum \rm limit
and \it quasi-classical \rm continuous energy absorbtion. First of
all, one may directly compare Eqs.(\ref{Elar}) and (\ref{P-w}):
calculating the average over one period in Eq.(\ref{P-w}) (which is
dominated by small $t\ll \omega^{-1}$ for $\Omega\ll\Omega_{0}$ and
by $\omega t\approx \pi/2$ for $\Omega\gg\Omega_{0} $) and replacing
$\Omega\rightarrow n\hbar\omega$ one indeed obtain the classical
result of Eq.(\ref{Elar}). An alternative way to see the crossover,
is to compute the moments of the number of absorbed/emitted photons
$\langle (n)^{m} \rangle$ exactly using Eq.(\ref{HG}). One obtains
\begin{eqnarray}
\label{mom} \langle (n)^{2} \rangle &=&\frac{1}{3}{\cal E}^2,\\
\nonumber \langle (n)^{4} \rangle &=&\frac{1}{3}{\cal
E}^2+\frac{9}{5}{\cal E}^4,\\ \nonumber \langle (n)^{6} \rangle &
=&\frac{1}{3}{\cal E}^2+9{\cal E}^4+\frac{225}{7}{\cal E}^6,\\
\nonumber \langle (n)^{8} \rangle &=&\frac{1}{3}{\cal
E}^2+\frac{189}{5}{\cal E}^4+450{\cal E}^6+1225{\cal E}^8.
\end{eqnarray}
The \it ultra-quantum \rm limit corresponds to all $\langle (n)^{2m}
\rangle=\frac{1}{3}{\cal E}^{2}$, i.e. keeping only the first term
in Eq.(\ref{mom}). On the other hand, the classical  distribution
[$3d$-case in Eq.(\ref{P-w})] leads to
moments $\langle (\Omega)^{2m} \rangle$ coinciding with the last
terms in  Eq.(\ref{mom}) upon the replacement $n\rightarrow
\Omega/\hbar\omega$. This shows again that multi-photon processes
drive the system at large intensities ${\cal E}\gg 1$ towards its
quasi-classical limit.

The table of moments Eq.(\ref{mom}) can in principle be extracted
from the smooth envelope of the distribution function
$f^{env}_t(E)$, using the standard techniques of the theory of
random walks~\cite{CTRW} to translate the properties of probability
kernel ${\cal P}_{\Omega}$ into a complete characterization of its
dynamics. It is clear that the finiteness of the second moment
$\langle (n)^{2} \rangle$ implies to zeroth order a standard
diffusive dynamics $f_{t}^{env}(E)\simeq {\rm
Erfc}\left[E/\sqrt{2D_E t}\right]/2$ ~\cite{YKK,Houches}, with
diffusion constant in energy space $D_E \equiv \frac{\omega^2
\langle (n)^2 \rangle}{\tau}= D [eE_0]^2$,
where $D=v_F^2 \tau/d$ is the diffusion constant in the real space.
Higher moments, such as $\langle (n)^{4} \rangle$, which in contrast
to the second moment do contain information about the \it
ultra-quantum \rm to \it quasi-classical \rm crossover, influence
higher order corrections. For example, up to first order in $\tau/t$
we obtain
\begin{eqnarray}\label{enve}
f^{env}_t(\omega) \simeq \frac{1}{2}{\rm
Erfc}\left[\frac{z}{\sqrt{2}}\right] + \frac{\lambda_4\; \tau
}{t}\frac{e^{-\frac{z^2}{2}}}{24 \sqrt{2\pi}}\;z(z^2-3) + \dots
\end{eqnarray}
where $z=\omega/\sqrt{D_E t}$ and $\lambda_4=\langle (n)^{4}
\rangle/\langle (n)^{2} \rangle^2-3$ is the kurtosis associated to
distribution $P_n$. The envelope $f_t^{env}(\omega)$ can in
principle be measured by a tunnelling experiment~\cite{Pothier}.
Note that although the term $\propto \lambda_4$ in Eq.(\ref{enve})
is a correction, it goes beyond the universal limit of RMT,
corresponding to $\tau \rightarrow 0$, and is determined by the
details of the semiclassical electron dynamics (e.g., smooth
disorder/anisotropic scattering, quantum isotropic scattering).

\section{Formal Derivation}~\label{quantitative}

{\setlength{\unitlength}{1cm}
\begin{figure}
\vglue -2cm \epsfxsize=0.8\hsize \vspace{0cm}
\hspace{-3cm}
\begin{picture}(4,4)
   \epsffile{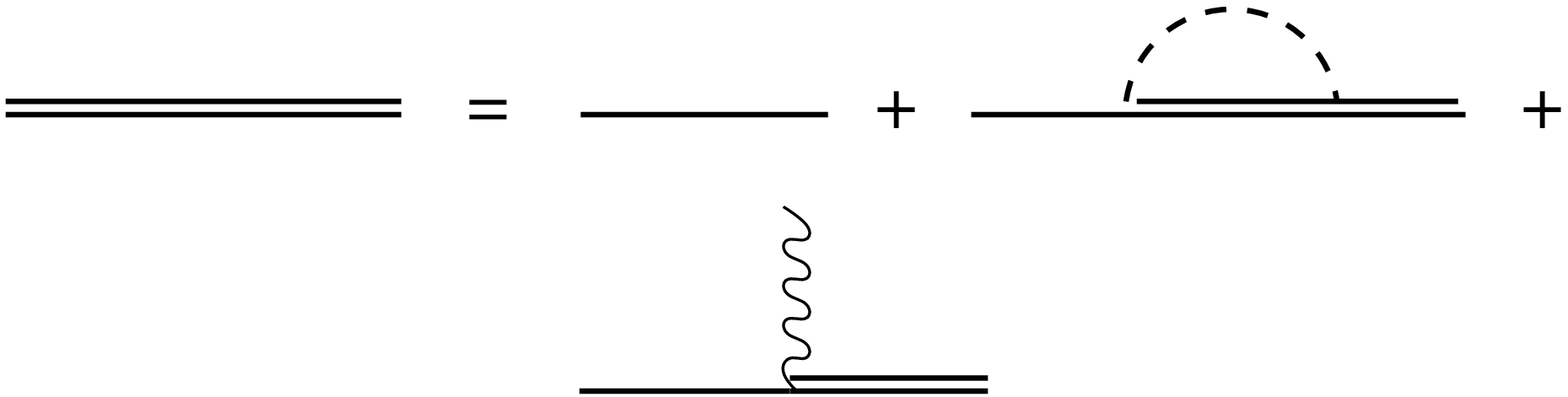}
\end{picture}\vspace{0cm}
\vspace{0cm} \caption{The self consistent Born approximation for the
retarded/advanced Green's functions Eq.(\ref{Born}) in its standard
diagrammatic representation~\cite{Houches,Abrikosov}. The
single/double lines represents the bare/full Green's function, wavy
lines correspond to the external drive $V(t)$, while the dashed line
is the disorder averaging.} \label{Fig2}
\end{figure}}

Let us now outline the formal derivation of the mapping of the
dynamics of the distribution function onto a continuous time random
walk in energy space, Eq.(\ref{KE}). We adopt a model of free
electrons in a Gaussian $\delta$-correlated static impurity
potential $U({\bf r})$ (which corresponds to isotropic scattering
amplitude) coupled to an external time-dependent vector potential
$\vec{A}(t)$, through $V(t)=- e\,\vec{v}\vec{A}(t)$. The Hamiltonian
takes the form
\begin{equation}
\label{Ham} \hat{H}=\frac{\hat{p}^2}{2m} + U({\bf r}) +V(t).
\end{equation}
where $\langle U(r)U(r')\rangle=1/(2\pi \nu \tau)\delta(r-r')$,
$\nu$ being the density of states at the Fermi level.

The distribution function $f_t(E)$ may be expressed as
$f_t(E)=\frac{1}{2}\left[1-\int d\eta e^{-i E\eta}
h_t(\eta)\right]$, where in turn $h_t(\eta)$ can be written in terms
of the disorder averaged Keldysh Green's function \cite{Keldysh,
Houches} as
\begin{eqnarray}
\label{h-G}
h_t(\eta)&=&\frac{i}{2\pi\nu}\int \frac{d{\bf r}}{V}
\langle G^{K}(t+\eta/2,t-\eta/2;{\bf r},{\bf r}) \rangle.
\end{eqnarray}

Let us now exploit the structure of the perturbative expansion of
$G^{K}$ in the time dependent perturbation $V(t)$. As shown in
detail in Ref.\onlinecite{Houches}, the noninteracting nature of
this problem makes it possible to identify two contributions to the
Keldish Green's functions, $\langle G^{K}\rangle=G^{K}_{0}+\delta
G^{K}$. The first
\begin{eqnarray}
G^{K}_0 &=& \int dt'' \langle G^{r}(t,t',r,r) \rangle h_0(t''-t')
\nonumber \\ && -h_0(t-t'') \langle G^{a}(t'',t',r,r) \rangle,
\end{eqnarray}
represents physically the unperturbed distribution function. Indeed,
calculating the disorder averaged retarded and advanced Green's
functions $\langle G^{r,a} \rangle$, within the self consistent Born
approximation [see Fig.~(\ref{Fig2})], one obtains
\begin{eqnarray}~\label{Born}
\langle G^{r,a}(t,t') \rangle_p &=& \mp i \theta(\pm t \mp
t')e^{-i\epsilon_p(t-t')}e^{\mp\frac{(t-t')}{2\tau}} \nonumber \\ &&
e^{-iv_F \vec{n} \cdot \int^{t}_{t'}dt'' e\vec{A}(t'')},
\end{eqnarray}
which immediately implies $\langle  G^{r,a}(t,t',r,r) \rangle= \mp i
\pi \nu \delta (t-t')$. Therefore,
$G^{K}_0(t+\eta/2,t-\eta/2,r,r)=-2\pi \nu i h_0(\eta)$.

{\setlength{\unitlength}{1cm}
\begin{figure}
\vglue 2cm \epsfxsize=0.8\hsize \vspace{0cm}
\hspace{-3cm}
\begin{picture}(4,4)
   \epsffile{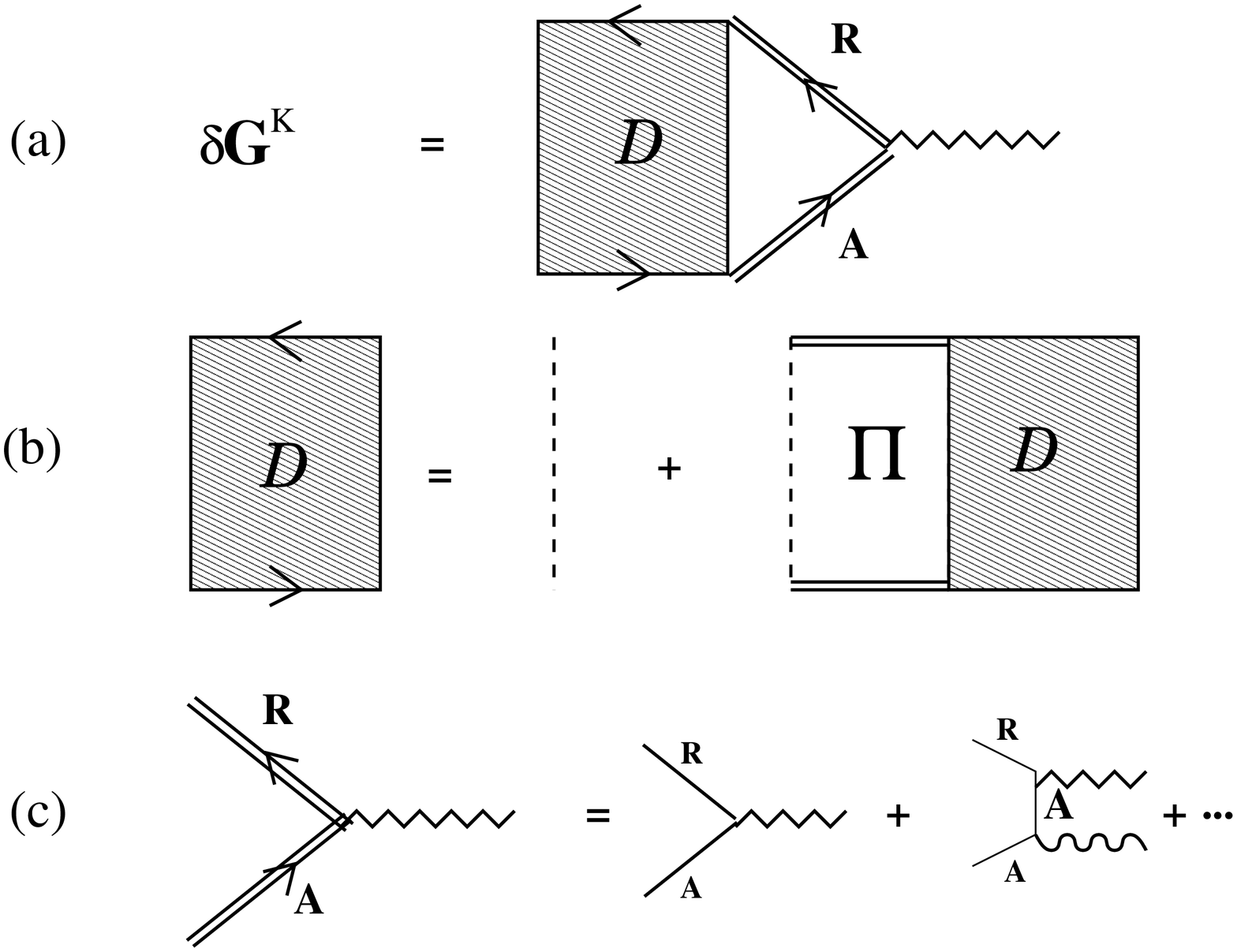}
\end{picture}\vspace{0cm}
\vspace{0cm} \caption{(a) The contribution to the Keldysh Green's
function $\delta G^K$. Eq.(\ref{diagram}) in its diagrammatic
representation to lowest order in the small parameter
$\delta/(v_F^{2}\tau A(t)^{2})$. The zig-zag line represents the
insertion of $h_0(t_1-t_2)[V(t_2)-V(t_1)]$. (b) The diagrammatic
representation of the diffuson appearing in $\delta G^K$. (c) The
perturbative expansion of the triangular vertex in the external
drive $V(t)$. } \label{Fig3}
\end{figure}}

The second contribution to $G^{K}$,
\begin{eqnarray}\label{anom}
 \delta G^{K} &=& \int dt_1 dt_2 dr_1
\langle G^{r}(t,t_1,r,r_1) h_0(t_1-t_2)[V(t_2) \nonumber \\ &&
-V(t_1)] G^{a}(t_2,t',r_1,r) \rangle,
\end{eqnarray}
describes energy absorbtion from the time dependent field.
Performing now the disorder average in Eq.(\ref{anom}) the product
of a retarded and advanced Green's function appearing in
Eq.(\ref{anom}) generates a diffusion propagator. More specifically,
$\delta G^{K}$ admits a diagrammatic representation in terms of a
loose diffuson [see Fig.(\ref{Fig3}-(a))], which formally amounts to
the equation
\begin{eqnarray}~\label{diagram}
\delta G^{K}=2 \pi i \nu  \int dt' dt'' {\cal D}_{\eta}(t,t') {\cal
L}_{\eta}(t',t'')\;h_0(\eta),
\end{eqnarray}
where, $D$ is the standard diffuson [see Fig.(\ref{Fig3}-(b))]
solution of the equation
\begin{eqnarray}
D^{-1} \otimes D_{\eta}&\equiv& D_{\eta}(t,t')-\int dt''
\Pi_{\eta}(t,t') D_{\eta}(t'',t') \nonumber \\
&=&\delta(t-t').
\end{eqnarray}
Neglecting interference effects, the kernel $\Pi$, as well as the
vertex ${\cal L}$ are given by
\begin{eqnarray}
\Pi_{\eta}(t,t')&=&\int d\eta'\,Tr\langle
G^{r}(t_{+},t'_{+})\rangle\,\langle
G^{a}(t'_{-},t_{-})\rangle/(2\pi\nu\tau) \nonumber \\
{\cal L}_{\eta}(t,t')&=&\int d\eta'\,Tr\langle
G^{r}(t_{+},t'_{+})\rangle\,\langle
G^{a}(t'_{-},t_{-})\rangle[V(t'_{-}))\nonumber \\
&-&V(t'_{+})]/(2\pi\nu i),\nonumber
\end{eqnarray}
where $t_{\pm}=t\pm\eta/2$ and $t'_{\pm}=t'\pm\eta'/2$, [see
Fig.(\ref{Fig3}(b)-(c))]. The $Tr$ symbol stands for the trace over
the coordinate indices; in particular it implies $\int d{\bf p}=\int
d\epsilon(p)\int d{\hat{n}}$ in the momentum representation where
the disorder averaged Green's functions $\langle G^{r,a}(t,t';{\bf
p})\rangle$ are diagonal. Preforming explicitly the trace, one
obtains
\begin{eqnarray}
\Pi_{\eta}(t,t')&=&\theta(t-t') \psi(t-t')\; p_{\eta}(t,t'),\\
{\cal L}_{\eta}(t,t')&=&\theta(t-t') \psi(t-t') \;
\partial_{t'}p_{\eta}(t,t'),
\end{eqnarray}
where $\psi(t)$ is given by Eq.(\ref{Poisson}) and
\begin{eqnarray}
p_{\eta}(t,t')&=&\int d\hat{n}\;\exp \bigg[i v_F \hat{n}\cdot
\hat{n}_x \bigg(\int_{t'+\eta/2}^{t+\eta/2}dt'' e A(t'')\nonumber
\\&-&\int_{t'-\eta/2}^{t-\eta/2} dt'' e A(t'')\bigg) \bigg].
\end{eqnarray}
Finally we may express $h_t(\eta)$ in terms of ${\cal D},{\cal L}$
as
\begin{eqnarray}\label{distribution}
h_t(\eta)= \left(1-\int dt' dt'' {\cal D}_{\eta}(t,t') {\cal
L}_{\eta}(t',t'')\right)\;h_0(\eta).
\end{eqnarray}

Let us now show that the distribution function is in the kernel of
the inverse diffusion propagator, i.e.
\begin{eqnarray}\label{kernel}
D^{-1}\otimes h_t(\eta)=0.
\end{eqnarray}
First of all notice that
\begin{eqnarray}
\tau \partial_{t'} \Pi_{\eta}(t,t')&=&
-\delta(t-t')+\Pi_{\eta}(t,t')+{\cal L}_{\eta}(t,t')\nonumber \\
&=& -\left[{\cal D}^{-1}\right]_{\eta}(t,t')+{\cal L}_{\eta}(t,t').
\end{eqnarray}
Now acting on Eq.(\ref{distribution}) with the operator $D^{-1}$ we
obtain
\begin{eqnarray}
\left[{\cal D}^{-1}\right]\otimes h_t(\eta)&=&\bigg[\int dt' {\cal
L}_{\eta}(t,t')-\tau \int dt' \partial_{t'} \Pi_{\eta}(t,t')
\nonumber \\ &-& \int dt' {\cal L}_{\eta}(t,t') \bigg]=0.
\end{eqnarray}
Notice at this point that Eq.(\ref{kernel}) is equivalent to the
recursion relation
\begin{eqnarray}
h_t(\eta)=\int dt' \; \Pi_{\eta}(t,t')\; h_{t'}(\eta).
\end{eqnarray}
Since $\int dt' \Pi_{\eta=0}(t,t')=1$, the latter may be
equivalently stated as
\begin{eqnarray}
f_t(\eta)&=&\int dt' \; \Pi_{\eta}(t,t')\; f_{t'}(\eta)\nonumber \\
&=& \int_0^{t} dt' \; {\cal P}_{\eta}(t,t')\; f_{t'}(\eta),
\end{eqnarray}
where $P_{\eta}(t,t')=\psi(t-t')p_{\eta}(t,t')$. The Fourier
Transform of this equation with respect to $\eta$ gives the
recursion relation defining the continuous time random walk,
Eq.(\ref{KE}).
 {\setlength{\unitlength}{1cm}
\begin{figure}
\vglue -3.4cm \epsfxsize=0.7\hsize \vspace{2.8cm}
\hspace{-3cm}
\begin{picture}(4,4)
   \epsffile{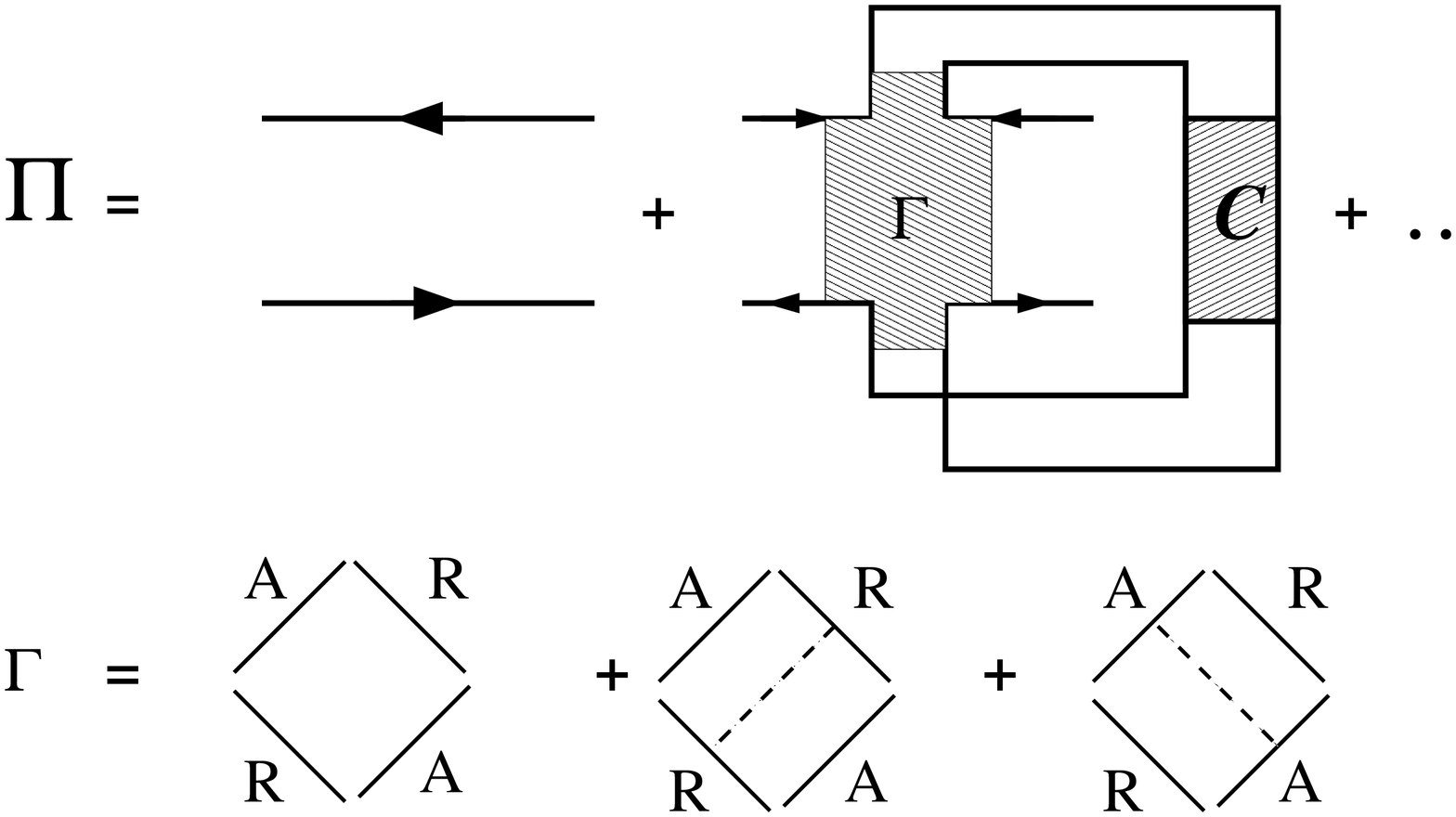}
\end{picture}\vspace{0cm}
\vspace{0cm} \caption{Loop expansion for $P_{\Omega}(t,t')$, the
insertion $C$ being a cooperon~\cite{Houches}. } \label{Fig4}
\end{figure}}

It is natural at this point to ask whether the formal description of
the time evolution of the distribution function as an effective
random walk may include interference/localization effects as well.
Indeed, performing a one-loop analysis with accuracy ${\cal E}^2$
one may shown~\cite{details} that the structure of Eq.(\ref{KE})
persists with a probability kernel schematically represented by the
diagrams in Fig.~(\ref{Fig4}). It is clear however that, in the
presence of interference, the Markovian nature of the random walk is
lost~\cite{notes}. In particular, upon time integration we obtain a
probability distribution $P_{\Omega}$ given by Eq.(\ref{RMT}) with
$p(t)\simeq\frac{{\cal E}^2}{3}\left(1-\sqrt{\frac{t}{t^*}}\right)$,
where the driving time $t$ is the time since the turning on of the
perturbation, $t^{*}=2\pi^3 D_E/(\Omega^2 \delta^2)$ is the
localization time in energy space~\cite{BSK}, and we neglected all
corrections independent of $t$. This result amounts to the weak
localization suppression of the absorbtion rate $W(t)/W_0 \simeq
1-\sqrt{t/t^*}$ due to weak dynamical localization~\cite{BSK}.

\section{Conclusions}~\label{conclusions}

In conclusion, we have studied the problem of energy
absorbtion/emission in a mesoscopic ring threaded by an oscillating
flux, focusing on the influence of multiphoton processes, and on the
multiphoton driven a crossover from ultra-quantum/low-intensity
limit, and quasi-classical/high intensity regime. We have shown that
the dynamics of the distribution function may be mapped onto a
continuous time random walk in energy space. Though in the present
paper we focused on the effect of a classical driving, recent
advances in the field of circuit QED~\cite{Vavi} strongly suggest
the possibility to investigate the role of single and multi-photon
processes in the case of quantum driving, an interesting problem
that remains a challenge for future work. ~\section{Acnowledgements}
We acknowledge useful discussions with D. Cohen, V. Falko, D.
Ivanov, M. Skvortsov, and V. Yudson.

\end{document}